\documentclass{article}
\usepackage[margin=1.4in]{geometry}
\usepackage[utf8]{inputenc}

\usepackage{amsmath}
\usepackage{amssymb}
\usepackage{amsthm}
\usepackage{amstext}

\usepackage{array}
\usepackage{graphicx}
\usepackage{wrapfig}
\usepackage{sidecap}
\usepackage{caption}
\usepackage[labelformat=simple]{subcaption}
\usepackage{footnote}
\makesavenoteenv{table}
\makesavenoteenv{tabular}

\theoremstyle{plain}
\newtheorem{lemma}{Lemma}[section]

\theoremstyle{definition}
\newtheorem{definition}[lemma]{Definition}
\theoremstyle{remark}

\usepackage{authblk}

\usepackage{biblatex}
\title{Analyzing Voting Power in Decentralized Governance: Who controls DAOs?}
\author{Robin Fritsch}
\author{Marino Müller}
\author{Roger Wattenhofer}
\affil{ETH Zürich\\ \texttt{\{rfritsch,muemarin,wattenhofer\}@ethz.ch}}
\date{}

\begin{document}

\maketitle

\begin{abstract}
We empirically study the state of three prominent DAO governance systems on the Ethereum blockchain: Compound, Uniswap and ENS.
In particular, we examine how the voting power is distributed in these systems.
Using a comprehensive dataset of all governance token holders, delegates, proposals and votes, we analyze who holds the voting rights and how they are used to influence governance decisions.

\end{abstract}

\textbf{Keywords:} DAO governance, liquid democracy

\section{Introduction}

Individuals with conflicting interests need to make joint decisions frequently in many parts of society: in political elections, at party conventions, or at shareholder meetings.
How can such communities govern themselves and reach joint decisions in the most effective and inclusive way?
Research on this question goes back centuries:
In 1884, Charles Dodgson (better known as Lewis Carroll, and author of Alice's Adventures in Wonderland) suggested a transitive voting system in which candidates can transfer received votes to other candidates~\cite{behrens2017liquid, carroll1884principles}.
Today, this concept is commonly referred to as \emph{liquid democracy}, and can be seen as a hybrid between direct and representative democracy.
It has lately been applied in practice by Germany's Pirate Party who used it for internal voting \cite{swierczek2011liquid, kling2015voting}.
Recently, a lot of experimentation and development in this area has been undertaken by communities around popular blockchain-based projects.
More and more such communities and stakeholder groups have formed so-called DAOs (decentralized autonomous organizations) in order to govern their projects in a decentralized manner.
Many of them have set up governance systems to make decisions about the future of their project.
Such decisions could concern changes to a protocol or the use of the treasury funds of a project.
With DAOs becoming more and more influential, so do their governance systems: Some DAOs already manage massive treasuries, the largest of them being Uniswap's \$4 billion treasury.\footnote{\url{https://openorgs.info/}}
And their governance decision can be highly consequential:
Uniswap's governance, for example, somewhat controversially decided to create a lobbying group called the ``DeFi Education Fund'' and allocate \$20 million of treasury funds to it in July 2021.\footnote{\url{https://app.uniswap.org/\#/vote/1/1}}
The novel designs of DAO governance systems and the large amount of power they hold make it compelling and relevant to study how these systems work and how decisions are reached.
Who holds the voting power in these systems and how do the actors influence the decisions?

The usage of applications built on top of blockchains has risen dramatically in the past years. Among the most popular are decentralized finance (DeFi) applications such as decentralized exchanges and lending protocols, as well as NFT projects.
In this paper, we study three of the most prominent governance systems on the Ethereum blockchain:
The first is Compound, a decentralized lending protocol \cite{leshner2019whitepaper}.
It was one of the first protocols to introduce a governance system and has been a role model for many other projects \cite{leshner2020governance}.
For instance, Uniswap, the second governance we examine, has adapted Compound's governance contracts.
Uniswap is a decentralized exchange and currently the largest decentralized finance project by market capitalization and treasury value.
Finally, we study ENS, the Ethereum Name Service, which is a vital part of Ethereum infrastructure with a large community that recently formed a governance.

\begin{wrapfigure}{r}{0.45\textwidth}
\includegraphics[width=0.45\textwidth, trim={0 0 0 2}, clip]{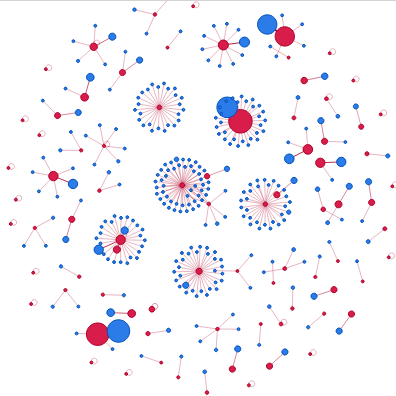}
\caption{A visualization of the delegation network with token holders (blue nodes) and delegates (red nodes) in Compound governance.}
\label{fig:network_comp}
\end{wrapfigure}

Most DAO governance processes, including the three we study, involve multiple steps, only the last of which takes place on the blockchain. For example, the Uniswap governance process first requires a temperature check and a consensus check, both of which are off-chain and require a certain threshold of yes-votes.
Only when these two thresholds are reached is a proposal put to a vote on-chain.
In this paper, we concentrate on the final on-chain votes where the actual decisions are made.

In order to facilitate voting in a decentralized and pseudonymous setting, DAOs issue governance tokens and distribute them among stakeholders. These tokens can initially be distributed in different ways: Uniswap for instance distributed their UNI tokens among the team, early investors and early users via an airdrop \cite{uniswap2020uni}.
Later, the tokens are freely tradable, making it possible to buy tokens and thereby voting power.
In this sense, governance tokens have similarities with shares in a company that come with voting rights at shareholder meetings.

The governance tokens can be used to vote on governance proposals: Each token counts as one vote.
For the voting, a simple form of liquid democracy \cite{paulin2020liquid_democracy} is used:
To be able to vote, a token holder must delegate their tokens to a \emph{delegate}.
Delegates can then vote on proposals with the amount of tokens delegated to them.
A visualization of the network of delegations for Compound is shown in Figure \ref{fig:network_comp}.
Note that both holders and delegates are addresses on the blockchain. In particular, multiple addresses can be controlled by the same person or organization.

In this paper, we analyze how voting power is distributed and used in the governance systems.
In the process, we evaluate how decentralized these systems actually are (since this is a major goal of blockchain-based applications).
Furthermore, we examine the structure of the delegation networks and study the voting behavior of different types of delegates.
Finally, we discuss why it is not always possible to determine the exact distribution of power since multiple addresses on the blockchain can be controlled by the same entity.

\section{Related Work}

A first attempt at creating a DAO was made with ``The DAO'' in the early days of the Ethereum blockchain in the year 2016 \cite{gemini2022daohack}.
However, due to a hack, The DAO never actually operated as intended.
One of the first governance systems around a blockchain-based application was introduced by the stablecoin protocol MakerDAO in 2018 \cite{maker2018governance}. A case study of MakerDAOs governance can be found in \cite{sun2022makerdao}.
Since Compound set up its governance system in early 2020 \cite{leshner2020governance}, many other projects have followed suit.
Some early analyses of the systems, mainly focussing on the ownership distribution of governance tokens of decentralized finance (DeFi) protocols, has been carried out in \cite{stroponiati2020, nadler2020defi_ownership, jensen2021decentralized}.
(For an overview of DeFi see \cite{werner2021defi_sok, aramonte2021defi}.)
In a recent, more detailed study, Barbereau et al.\ consider the governance systems of nine DeFi protocols: Uniswap, Aave, MakerDAO, Compound, SushiSwap, Synthetix, Yearn Finance, 0x, and UMA \cite{barbereau2022decentralised}.
In their work, they focus mainly on token distribution and voter turnout.
The analysis in our paper goes deeper; we also examine voting behavior and delegation network structure.


Voting systems similar to those currently used by many DAOs have been studied for a long time. They are often referred to as delegative democracy \cite{ford2002delegative} or liquid democracy \cite{behrens2017liquid, paulin2020liquid_democracy}.
The governance systems we study use slightly simpler versions of these concepts with a main difference being that they only allow for voting rights to be delegated once from a token holder to a delegate. In liquid democracy, it is often possible for a delegate to further (recursively) delegate the tokens they received to another delegate.
The use of liquid democracy in practice has been studied in a case study of the German Pirate Party's internal voting system \cite{swierczek2011liquid, kling2015voting}.

\section{Overview of the Dataset}
\label{sec:dataset}

We collect data on the token holders, delegates and proposals for the Compound, Uniswap and ENS governance systems from the Ethereum blockchain.
Table~\ref{tbl:dataset} shows an overview of the state of the governance systems on 1 March 2022.
We consider the time period between blocks 9.000.000 (25 November 2019) and 14.300.000 (1 March 2022).
The data was gathered using the indexing protocol The Graph\footnote{\url{https://thegraph.com/hosted-service/subgraph/arr00/compound-governance-2},\\ \url{https://thegraph.com/hosted-service/subgraph/arr00/uniswap-governance-v2},\\ \url{https://thegraph.com/hosted-service/subgraph/ianlapham/ens-governance}} to query data from the Ethereum blockchain.

\begin{table}[ht]
\centering
\begin{tabular}{ |l|>{\centering\arraybackslash}m{2cm}|>{\centering\arraybackslash}m{2cm}|>{\centering\arraybackslash}m{2cm}|>{\centering\arraybackslash}m{2cm}| } 
 \hline
  & Holders & Delegates & Proposals & Total Votes \\
 \hline
 Compound & 188.198 & 1.766 & 84 & 3.431 \\ 
 Uniswap & 306.580 & 5.784 & 10 & 1.660 \\ 
 ENS & 57.511 & 11.597 & -\footnote{Up until 1 March 2022, two proposals were voted on by ENS governance, however these were not part of our dataset.} & - \\ 
 \hline
\end{tabular}
\caption{Governance systems.}
\label{tbl:dataset}
\end{table}

\begin{figure}
\begin{subfigure}{0.50\textwidth}
\centering
\includegraphics[width=0.9\linewidth, trim={0 0 0 2}, clip]{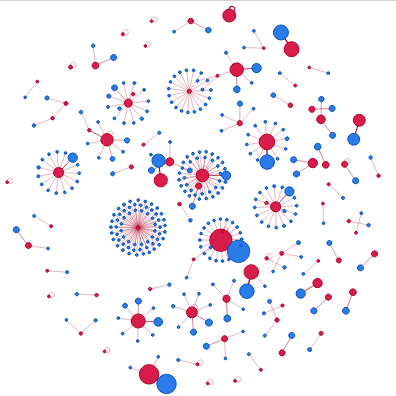}
\caption{Uniswap}
\label{fig:network_uni}
\end{subfigure}%
\begin{subfigure}{0.50\textwidth}
\centering
\includegraphics[width=0.9\linewidth, trim={0 0 0 2}, clip]{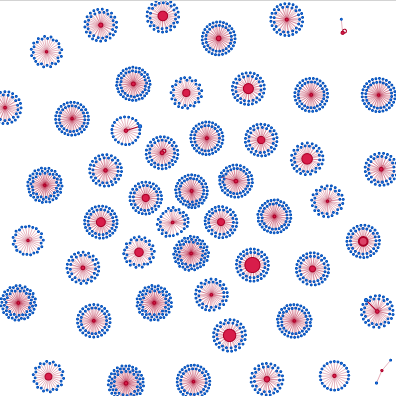}
\caption{ENS}
\label{fig:network_ens}
\end{subfigure}
\caption{Delegation Networks: Blue nodes show holders and red nodes delegates. The sized of the nodes are proportional to the amount of tokens held or represented. An edge indicates a delegation.}
\label{fig:network}
\end{figure}

The delegation networks of the governance systems are visualized in Figures \ref{fig:network_comp} and \ref{fig:network} using the Python library NetworkX~\cite{networkx}. An edge between a holder (blue node) and a delegate (red node) indicates a delegation. The sizes of the nodes are proportional to the amount of tokens held or represented (with a minimum size for small holders and delegates).

We already see a clear difference between Compound and Uniswap on the one side and ENS on the other side: For ENS, delegates tend to have a large number of holders of similar size delegating to them. For Compound and Uniswap on the other hand, many delegates receive most of their voting power from a single large token holder delegating to them. Note that in such cases it is possible that the holder and delegate addresses are both controlled by the same entity (and the delegate address was set up for voting purposes only).
We investigate the difference between the governance systems in more detail in Section \ref{sec:del_class}.

\section{Distribution of Voting Power}

In the following, we study who holds the power in the governance systems.
We examine this in three steps: 1) How are the voting rights distributed? 2) How frequently are the voting rights executed? 3) How much do the executed votes influence the final governance decision?

\subsection{Distribution of Voting Rights}

\begin{figure}[ht]
    \centering
    \includegraphics[width=0.6\textwidth]{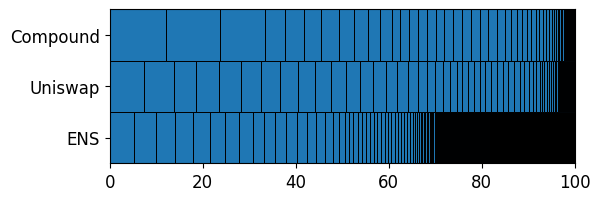}
    \caption{Distribution of voting rights among delegates.}
    \label{fig:delegates_dist}
\end{figure}

As a first step, we examine how voting rights are distributed in the three governance systems. The distributions of delegated tokens among delegates are shown in Figure \ref{fig:delegates_dist}.
From the visualization, we already see that the voting rights are concentrated among a relatively small number of delegates.
To quantify this, we consider two inequality measures: the Gini coefficient and the Nakamoto coefficient.

\begin{figure}[ht]
\centering
\begin{subfigure}{.33\textwidth}
  \centering
  \includegraphics[width=\linewidth]{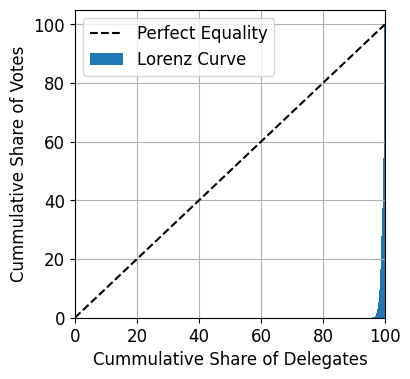}
  \caption{Compound}
  \label{fig:lorenz_curve_comp}
\end{subfigure}%
\begin{subfigure}{.33\textwidth}
  \centering
  \includegraphics[width=\linewidth]{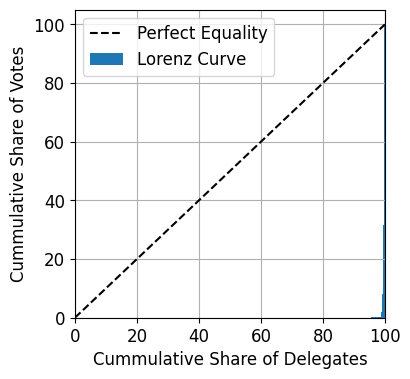}
  \caption{Uniswap}
  \label{fig:lorenz_curve_uni}
\end{subfigure}%
\begin{subfigure}{.33\textwidth}
  \centering
  \includegraphics[width=\linewidth]{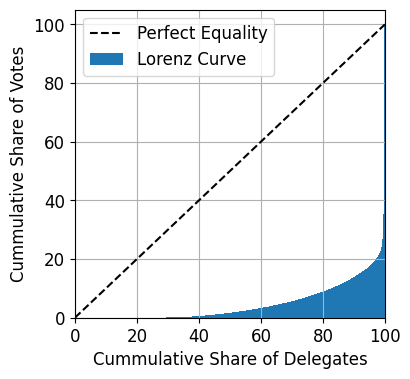}
  \caption{ENS}
  \label{fig:lorenz_curve_ens}
\end{subfigure}
\caption{Lorenz curves of the token distributions among delegates.}
\label{fig:lorenz_curves}
\end{figure}

The Gini coefficient \cite{gini} is the most commonly used index to measure inequality and ranges from 0 (perfect equality) to 1 (maximum inequality).
It measures area between the Lorenz curve of a given distribution and the curve of perfect equality \cite{gini}.
The Lorenz curves of the distributions of tokens among delegates are shown in Figure \ref{fig:lorenz_curves}.
For all of this section, we consider the distributions of tokens at block 14.300.000 which occurred on 1 March 2022.
We clearly see that ENS' token distribution is less unequal that the other two.
The precise values of the Gini coefficient can be found in Table \ref{tbl:gini_nakamoto}.
They show that the inequality in the distribution of voting rights is very high in all three cases, the most extreme being Compound and Uniswap with Ginis around 0.99.
In particular, the inequality in the general distribution of wealth in the world is a lot lower:
The Gini coefficient of the wealth distribution in 2020 is estimated to be 0.850 in the US, 0.814 in Europe and 0.704 in China  \cite{cs2021wealth}.

We also see that the distribution of tokens among delegates is slightly more equal than the distribution among token holders. Furthermore, considering only delegates who voted at least once gives a slightly more equal distribution once again.

The Nakamoto coefficient \cite{srinivasan2017quantifying} is a measure of how decentralized a system is. Applied to our setting, it counts the minimum number of delegates that together hold more than 50\% of the voting power.
The Nakamoto coefficients (see Table \ref{tbl:gini_nakamoto}) reveal an extreme centralization in all three governance systems:
In the case of Compound, only 8 delegates can dictate any governance action using their majority. For Uniswap this number is 11, for ENS it is 18.

\begin{table}[ht]
\centering
\begin{tabular}{ |l|>{\centering\arraybackslash}m{2cm}|>{\centering\arraybackslash}m{2cm}|>{\centering\arraybackslash}m{2cm}| } 
 \hline
  & Compound & Uniswap & ENS \\
 \hline
 Gini Holders & 0.998 & 0.996 & 0.977 \\ 
 Gini Delegates & 0.987 & 0.995 & 0.908 \\ 
 Gini Delegates Voted & 0.964 & 0.971 & - \\ 
 Nakamoto Delegates & 8 & 11 & 18 \\
 \hline
\end{tabular}
\caption{Gini and Nakamoto coefficients of the token distributions among holders, delegates and delegates who voted at least once. All numbers are rounded to 3 digits after the comma.}
\label{tbl:gini_nakamoto}
\end{table}

\begin{figure}
    \begin{subfigure}{.5\textwidth}
        \centering
        \includegraphics[width=\linewidth]{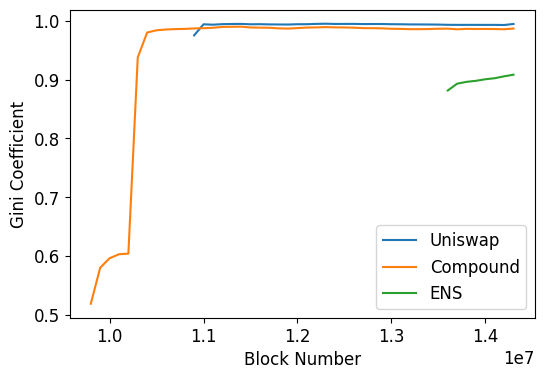}
        \caption{}
        \label{fig:gini}
    \end{subfigure}%
    \begin{subfigure}{.5\textwidth}
        \centering
        \includegraphics[width=\linewidth]{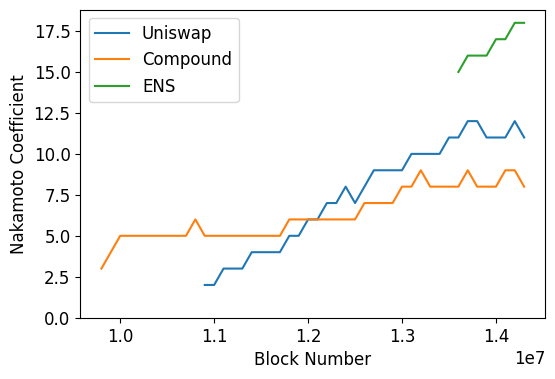}
        \caption{}
        \label{fig:nakamoto}
    \end{subfigure}
    \caption{Gini and Nakamoto coefficients over time.}
    \label{fig:gini_nakamoto}
\end{figure}

When we look at the Gini and Nakamoto coefficients over time (see Figure \ref{fig:gini_nakamoto}), we see that the extreme inequality in the voting power distributions has persisted for a long time and even seems to be rising.
On the other hand, the rising Nakamoto coefficients also indicate that the systems are becoming more decentralized over time.

\subsection{Governance Participation}

A second crucial aspect in analyzing voting power is considering how often delegates participate in voting, i.e.\ how often delegates use their voting power.

\begin{figure}
\centering
\begin{subfigure}{.5\textwidth}
  \centering
  \includegraphics[width=0.95\linewidth]{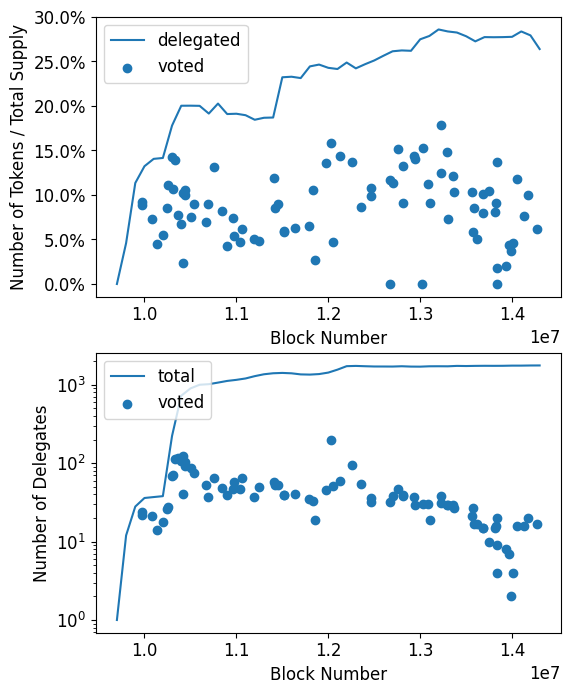}
  \caption{Compound}
  \label{fig:gov_participation_comp}
\end{subfigure}%
\begin{subfigure}{.5\textwidth}
  \centering
  \includegraphics[width=0.95\linewidth]{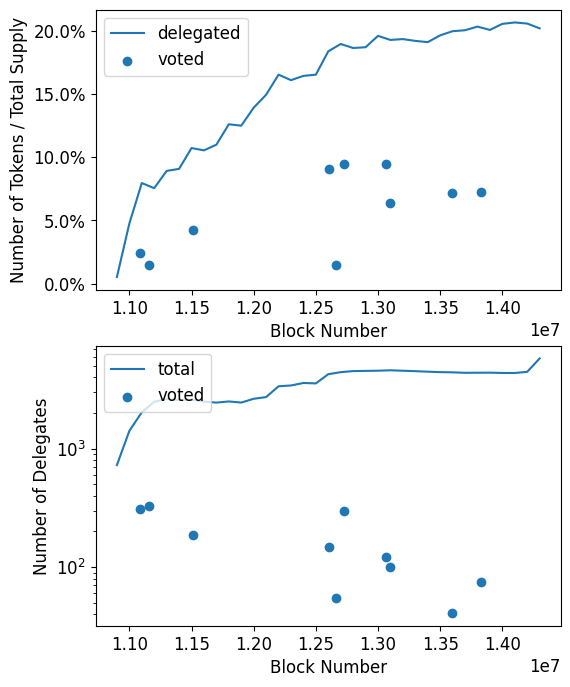}
  \caption{Uniswap}
  \label{fig:gov_participation_uni}
\end{subfigure}
\caption{Governance Participation: The upper figures show the amount of tokens delegated and how many of them were used to vote in each proposal. The lower figures plot the number of delegates in total and how many of them voted.}
\label{fig:gov_participation}
\end{figure}

The upper plots in Figure \ref{fig:gov_participation} show the fraction of the total supply of governance tokens which is delegated (the total supply is 10 million for Compound and 1 billion for Uniswap). Since delegating is a prerequisite of voting, this is the maximum amount of tokens that could have participated in governance at that time. Furthermore, the figures show how many of these tokens were actually used to vote for each proposal (each dot represents a proposal).

Note that only about two thirds of the total supply of COMP and UNI tokens are actually circulating at the end of our observation period according to CoinMarketCap\footnote{\url{https://coinmarketcap.com/}}. The remaining amounts will come into circulation in the future. Since only circulating tokens are available for delegating, a delegation rate of about 20\% for Uniswap translates to 30\% of all circulating tokens.
As a consequence, 15\% of circulating governance tokens can currently take control of Uniswap's governance.
Note that circulating supply has increased over time which could in part explain the increase in delegated tokens.

The figures show that, on average, in both governance systems less than 10\% of total tokens (or 15\% of circulating tokens) participate in votes on proposals.
Hence, already 7.5\% of circulating tokens would have sufficed to win an average vote in the past.

To give a comparison, participation in voting at shareholder meetings of traditional companies (which arguably share some similarities at least with Uniswap and Compound governance) is a lot higher:  According to \cite{zachariadis2020free}, the participation rate among shareholders of US companies is about 70\%.

The lower plots in Figure \ref{fig:gov_participation} show the number of delegates in total and how many of them participated in each vote.
We find that only a very small number of delegates actually participates in voting (note the logarithmic $y$-axis). This again shows how the centralized these governance systems are.
For Compound the number has been falling and the average is currently around 20 while it has lately been below 100 for Uniswap.

\begin{figure}
\centering
\begin{subfigure}{.5\textwidth}
  \centering
  \includegraphics[width=0.95\linewidth]{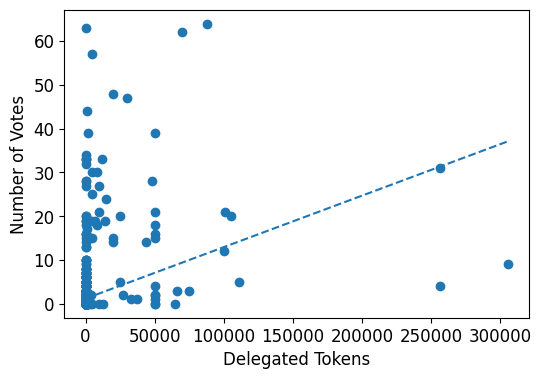}
  \caption{Compound}
  \label{fig:del_votes_comp}
\end{subfigure}%
\begin{subfigure}{.5\textwidth}
  \centering
  \includegraphics[width=0.95\linewidth]{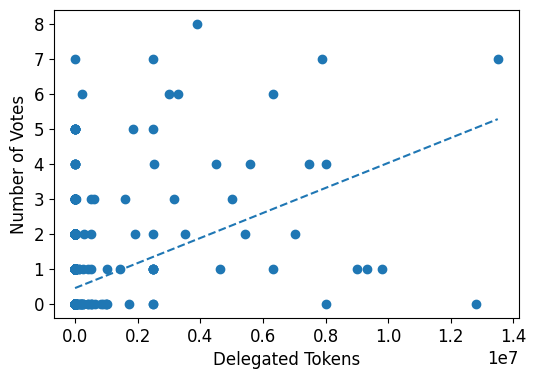}
  \caption{Uniswap}
  \label{fig:del_votes_uni}
\end{subfigure}
\caption{Delegate governance participation.}
\label{fig:del_votes}
\end{figure}

Lastly, we take a look at how the participation rate of delegates depends on the number of tokens delegated to them.
Figure \ref{fig:del_votes} plots how often delegates vote against the number of delegated tokens.
Each dot represents a delegate, and the dashed line shows the regression line.
We observe that delegates with more tokens delegated to them tend to vote more often. 
A notable exception is the second largest Uniswap delegate, who has been delegated almost 12.8 million UNI tokens (from a single address) worth more than \$13 million on 1 March 2022 but has never voted in a governance decision.

\subsection{Potential and Exercised Voting Power}

As the final step, we examine how the votes that are cast actually influence the outcome of the governance decisions.
It is one thing to have the potential to change a vote and another to actually do it.
We measure this using the following two metrics based on Kling et al.~\cite{kling2015voting} which are designed to quantify the potential and exercised voting power of delegates.

\begin{definition}[Potential Voting Power]
We measure the potential voting power of a delegate by counting how often the delegate could have changed the outcome of a proposal vote by changing their own vote.
Formally, the potential power $p_{pot}\in\{0,1\}$ of a delegate who participated in a proposal vote is defined as 
\begin{equation*}
    p_d^{pot} =
        \begin{cases}
           1  & \text{if } v_d > \frac{1}{2}\left(V^{(y)} + V^{(n)}\right)-V^{(y)}+\delta_d v_d > 0  \\
           0  & \text{else}
        \end{cases}
\end{equation*}
where $V^{(y)}$ and $V^{(n)}$ are the number of Yes- and No-votes on the proposal, respectively. Furthermore, $v_d$ denotes delegate $d$'s number of votes and $\delta_d\in\{0,1\}$ indicates whether delegate $d$ voted Yes (1) or No (0) on the proposal.

The overall potential power of a delegate is then defined as the mean of $p_d^{pot}$ over all proposals.
\end{definition}

\begin{definition}[Exercised Voting Power]
The exercised voting power measures how often the vote of a delegate actually changed the outcome of a proposal vote, i.e. how often the result would have been different without the delegate's votes.
It is defined as
\begin{equation*}
    p_d^{ex} =
        \begin{cases}
           1  & \text{if } \left(\frac{V^{(y)} - \delta_d v_d}{V^{(y)} + V^{(n)} - v_d}>\frac{1}{2}\right) \neq 
           \left(\frac{V^{(y)}}{V^{(y)} + V^{(n)}}>\frac{1}{2}\right)  \\
           0  & \text{else.}
        \end{cases}
\end{equation*}
Again, the overall exercised power is then defined as the mean over all $p_d^{ex}$.
\end{definition}

\begin{figure}
\centering
\begin{subfigure}{.5\textwidth}
  \centering
  \includegraphics[width=0.95\linewidth]{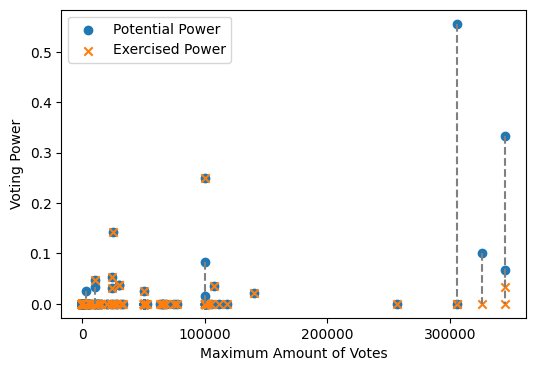}
  \caption{Compound}
  \label{fig:voting_power_comp}
\end{subfigure}%
\begin{subfigure}{.5\textwidth}
  \centering
  \includegraphics[width=0.95\linewidth]{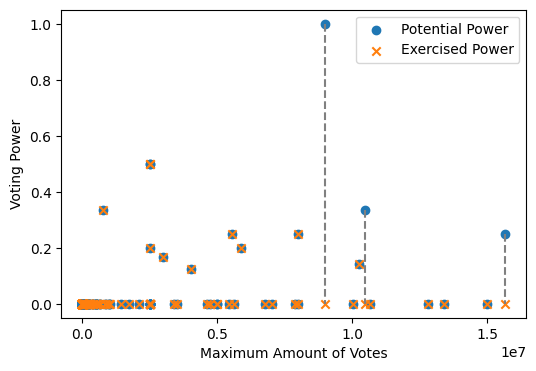}
  \caption{Uniswap}
  \label{fig:voting_power_uni}
\end{subfigure}
\caption{Potential and exercised voting power.}
\label{fig:voting_power}
\end{figure}

Figure \ref{fig:voting_power_comp} shows the potential and exercised voting power of delegates in Compound governance system.
We see that most of the potential voting power is shared among four of the largest delegates. One of them could have changed the outcome of 5 of the 9 proposals they voted one. 
This is again evidence of how much power lies in the hands of very few individuals.
However, we also see that these individuals do not exercise their power a lot: Their exercised voting is far smaller than their potential power.

The same is true for Uniswap (see Figure \ref{fig:voting_power_uni}): Large delegates only exercise their voting power very rarely.
We look into this in more detail in Section \ref{subsec:large_small} when we compare how large and small delegates vote and find that large delegates often vote in line with small delegates.

Note that most potential and exercised voting power of small delegates that can be seen in the figures stems from two proposals with very close outcomes (which means that they could have been influenced by many delegates):
Proposal 16 for Compound\footnote{\url{https://compound.finance/governance/proposals/16}} and Proposal 7 for Uniswap\footnote{\url{https://app.uniswap.org/#/vote/1/2}}.

\section{Delegate Classification and Voting Behavior}
\label{sec:del_class}

In this section, we investigate the voting behavior of different types of delegates. 
We have seen in Section \ref{sec:dataset} that ENS seems to have a different structure in its delegation network.
In the following, we will quantify these differences.
To that end, we separate the delegates into two groups:
If a delegate receives at least 50\% of their voting power from a single holder we classify them as a \emph{single holder delegate}.
The remaining delegates are classified as \emph{community delegates}.

\subsection{Single Holder vs. Community Delegates}
\label{subsec:comm_single}

\begin{figure}
    \centering
    \includegraphics[width=0.6\textwidth]{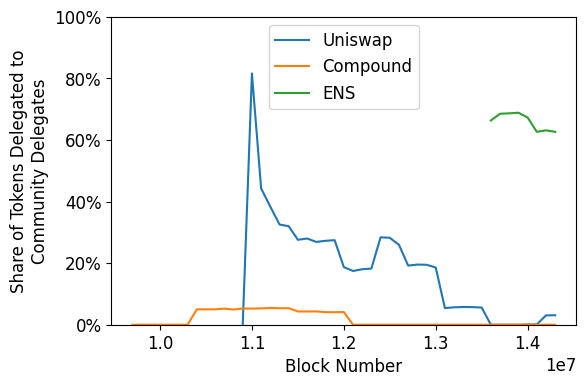}
    \caption{Share of delegated tokens controlled by community delegates (i.e.\ delegates who receive less than 50\% of their votes from a single address).}
    \label{fig:community_delegates}
\end{figure}

Figure \ref{fig:community_delegates} shows the share of delegated tokens that is controlled by community delegates.
The difference becomes clear right away: For ENS, about 60\% of voting rights are in the hands of community delegates while it is less than 10\% for Compound and Uniswap.
While the fraction of voting rights held by community delegates has always been low for Compound, the community delegates actually had more influence in the early stages of Uniswap governance.

This observation shows that delegations are currently not being used to a great extent in Compound and Uniswap governance. Most voting power lies in the hands of delegates who receive most of their voting power from a single token holder. It is possible that in many such cases, the two addresses are controlled by the same entity.
In this regard, Compound and Uniswap governance show similarities to shareholder meetings where large investors represent their interests.
ENS on the other hand, shows more similarities to a decentralized community.
 
Next we compare the voting behavior of community delegates and single holder delegates.
We find that both groups vote similarly:
They only voted differently on 3 Compound proposals and a single Uniswap proposal.

\subsection{Large vs. Small Delegates}
\label{subsec:large_small}

We compare the voting behavior of large and small delegates. The biggest delegates that together hold at least 90\% of voting rights at the time of a proposal are classified as \emph{large delegates}. The remaining delegates are \emph{small delegates}.

It is clear that large delegates, defined in this way, hold all power in the governance system and can decide any vote in their favor.
But how often do they actually force against the minority of small delegates.

We find that this only rarely happens:
In Compound governance, large delegates only voted differently from small ones on 6 out of the 84 proposals.
For Uniswap the picture is similar:
Here large and small delegates only disagreed on a single proposal, and that only very narrowly: On proposal 7, large delegates were 50.5\% in favor, while small delegates opposed it (34.5\% were in favor).

\section{Address Clustering}

In general, multiple addresses (holder as well as delegate addresses) can be controlled by the same entity.
This makes it difficult to determine who actually holds the power and can make the governance systems seem more decentralized than they actually are.

Moreover, the situations of a single entity controlling multiple addresses is actually favored by the design of the delegation rules of the governance systems:
It is only possible for a token holder to delegate all of their tokens to a single address. If a token holder would like to delegate to multiple delegates, they need to transfer their tokens to multiple addresses and delegate from there.

An example of an entity requiring multiple addresses, could be a delegation program of a large token holder, such as that of the venture capital firm a16z~\cite{a16z_delegation_program}.

We try to detect such situations by considering transfers of governance tokens between addresses: If tokens are sent from one address to another, it is likely that the two addresses are controlled by the same entity.
Examining these transfers, creates a network of addresses with links between them.
In this network, we look for connected components.
During the process, we exclude special addresses such as exchange addresses.
Using this method, we find a number of groups of addresses of Compound and Uniswap token holders which are potentially controlled by a single entity.

\begin{figure}
\centering
\begin{subfigure}{.5\textwidth}
  \centering
  \includegraphics[width=0.90\linewidth]{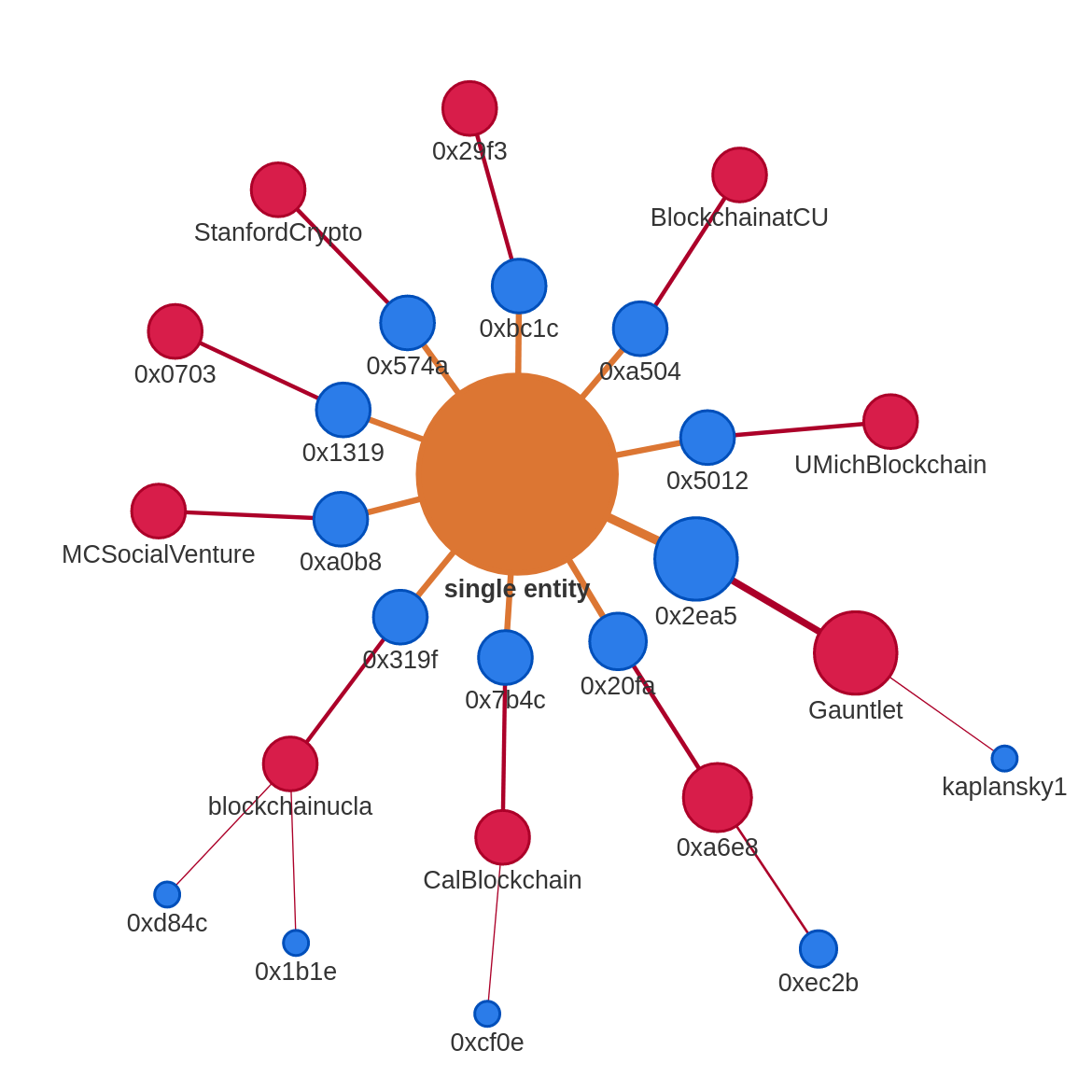}
  \caption{Compound}
  \label{fig:single_entity_comp}
\end{subfigure}%
\begin{subfigure}{.5\textwidth}
  \centering
  \includegraphics[width=0.90\linewidth]{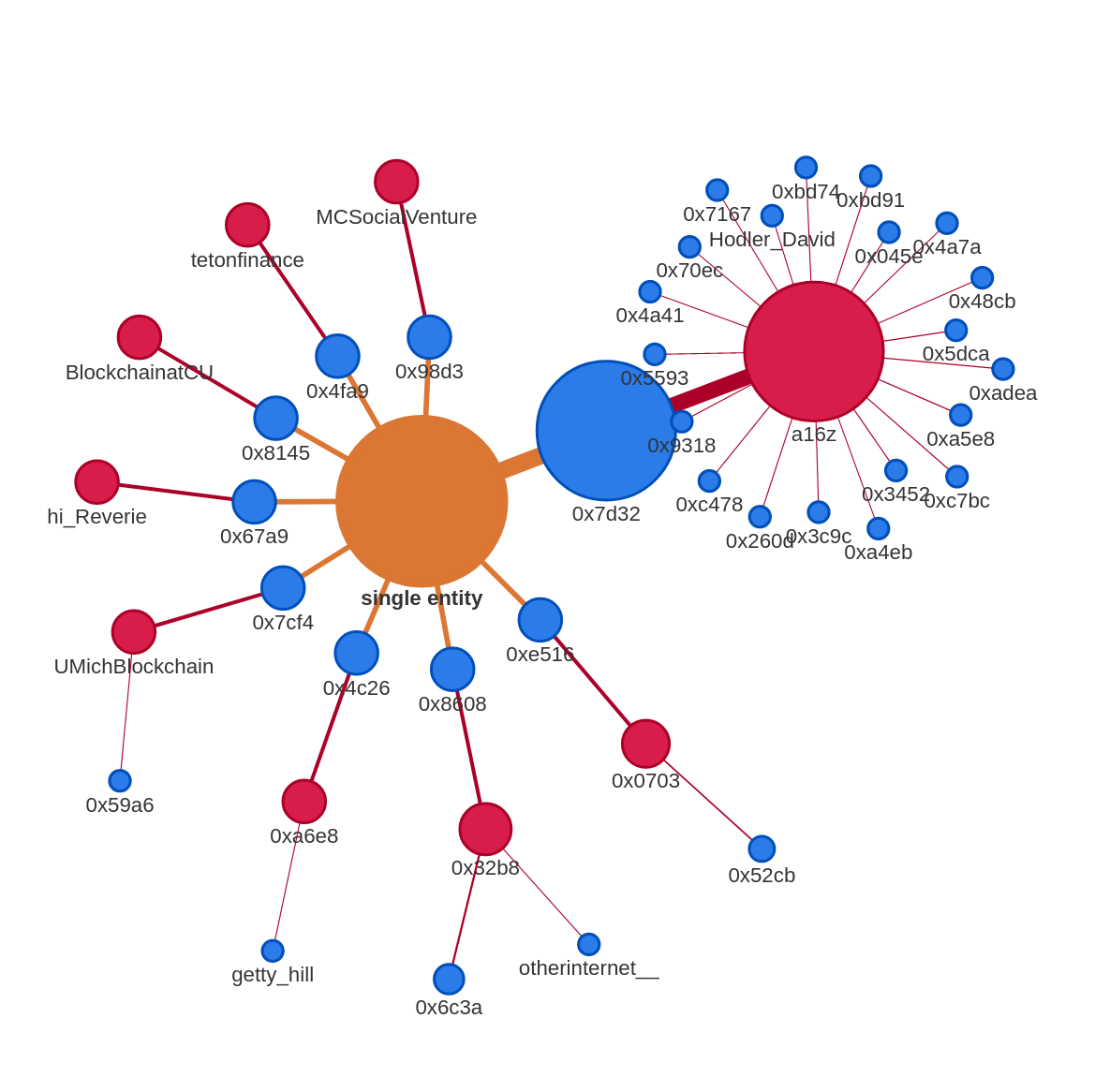}
  \caption{Uniswap}
  \label{fig:single_entity_uni}
\end{subfigure}
\caption{Addresses likely controlled by a single entity based on token transfers.}
\label{fig:single_entity}
\end{figure}

Figure \ref{fig:single_entity} shows two examples of such groups.
Possibly these addresses are part of a16z's delegation program~\cite{a16z_delegation_program}. 
Of course, this method is not bullet-proof and we do not claim that the addresses shown in the figure are for certain controlled by the same entity.
With these examples, we would like to demonstrate that blockchain-based governance systems are not always fully transparent, and power in them can be more centralized than it seems.

\section{Conclusion}

Our analysis shows that the power in the governance systems of the DeFi protocols Compound and Uniswap is extremely centralized and controlled by a very small number of addresses.
Their governance systems show similarities with shareholder meetings, where a small number of large investors represent their interests. (ENS governance, on the other hand, more resembles of a decentralized community.)
However, we also see that large delegates do not often use their power and mostly decide in the same way as the larger community, i.e. smaller delegates.
We state these observations without judgement. Nonetheless, they do illustrate the challenges of building a truly decentralized  governance systems.

As the importance of DAOs grows over time, it will become increasingly interesting to continue researching along the lines of this paper.
It will be possible to study a larger number of DAO governance systems and more proposal votes in the future.

\newpage

\printbibliography

\end{document}